\documentclass[12pt]{article}
\usepackage{latexsym}
\usepackage{graphics}
\usepackage{epsfig}
\usepackage{epstopdf}
\usepackage{amsmath}
\usepackage{cite}
\usepackage{hyperref}
\usepackage{amssymb}

\begin{document}

\begin{center}
\textbf{\Large Properties of impact events  in the model of forced impacting oscillator: experimental and numerical investigations} \vspace{0.5 cm}

Sergii Skurativskyi$^\dagger$ 
\footnote{e-mail:
\url{skurserg@gmail.com}},
 Grzegorz Kudra$^\ddagger$ \footnote{e-mail:
\url{grzegorz.kudra@p.lodz.pl}},
Grzegorz Wasilewski$^\ddagger$ \footnote{e-mail:
\url{grzegorz.wasilewski@p.lodz.pl}},
Jan Awrejcewicz$^\ddagger$ \footnote{e-mail:
\url{jan.awrejcewicz@p.lodz.pl}}

 \vspace{0.5 cm}

$^\dagger$Subbotin Institute of Geophysics, NAS of Ukraine, Kyiv, Ukraine

$^\ddagger$Lodz University of Technology, Department of Automation, Biomechanics and Mechatronics,
Lodz, Poland
\end{center}

\begin{quote} \textbf{Abstract. }{\small 
The paper deals with the studies of forced impacting oscillator when are taken  into account the dry and viscous resistance, as well as the generalized  Hertz  contact law during an impact. The numerical treatments of mathematical model are accompanied with the validations on the base of experimental rig. To study the solutions of the  mathematical model, we construct the sequences of impacts, when  the system is evolved in periodic and chaotic modes. The statistical properties of chaotic impact events are considered in more details. In particular, we analyze  the successive iterations of impact map, autocorrelation function and coefficient of variation for the impact  train, the histograms for the inter-impact  intervals  and  values of obstacle penetrations. It is revealed that the impact sequence is stationary but non-Poissonian and contains temporal scales which do not relate to the external stimulus.  This sequence  can be described by a bimodal distribution. These findings are confirmed by the analysis of  experimental data. 
}
\end{quote}

\begin{quote} \textbf{Keyword: }{\small 
 forced oscillator; bifurcations;
impact; spike statistics; chaos
}
\end{quote}

\vspace{0.5 cm}

\section*{Introduction}

Impacts phenomena are diverse and widely encountered in mechanical systems. The studies of such systems can be  realized within the framework of non-smooth dynamical  models \cite{NonSmooth2004,Jan2003} where the limiters of motion are introduced. Due to nonsmoothness, the consideration of these  mathematical models can lead  to the class of strongly nonlinear models possessing  special types of behavior \cite{Jan2003, ChongIJNM, Witelski2014,Mikhlin98}. 

As has been shown in a wide range of   works   \cite{Shaw83, Peterka92,Okolewski2014,Okolewski2017}, impacting systems subjected to external excitation  possess harmonic, subharmonic,  chaotic and other complicated motions. The bifurcation phenomena  both  specific  (i.e., various types of grazing bifurcations \cite{Grazing1994,Balachandran2008})  and  are inherent in smooth systems \cite{Isomaki85}, take place when the control parameters were varied.
To date, thus, the general foundations of one-degree-of-freedom impacting systems, their theoretical description and experimental   validation, were developed essentially. 

  But the peculiarities of impacting system  when multiperiodic or chaotic regimes occur are still 
  poorly investigated. To deeper understand the nature of complicated regimes, the analyses of the sets of discrete events extracted from the system's dynamics can be useful. Events, in the terms of  neuroscientists \cite{Gerstner2014} also named as spikes,   
can  be  identified  as  a 
abrupt  change of
system's  variable \cite{Racicot97}.   To generate the spike sequences, as a rule, the threshold-crossing  and integrate-and-fire  techniques \cite{AnishchenkoPRE2000, AnishchenkoPRE2001} are employed. In the case of impacting system, the natural threshold is present, namely the limiter.  When the system's trajectory  crosses the level defined by limiter, the temporal moments of impacts form the impacting map \cite{Lee2005} statistical properties of which, especially  in chaotic modes, are extremely interesting.   

It should be noted that the impact sequences can be associated with the Poincare sections which in the case of harmonic external loading are especially simply defined. Among the advantages of impact sequences \cite{Slade97}, let us note the situation when the loading is unharmonic. Then the Poincare section technique for nonautonomous models is required some modifications, whereas the studies of impacts can be carried out in the same manner. Moreover, inter-impact intervals  allow us to assess the dynamical and geometrical properties of chaotic attractors \cite{AnishchenkoPRE1998}. 

Impact events (as well as the displacement or velocity of a cart) is part of information directly produced by the oscillating system. The searching for hidden regularities in this information motivates us to treat the discrete event sequences and develop proper tool for these purposes. 

Thus,  the   paper is organized as follows. In section \ref{skur:secOne} we describe the experimental rig and corresponding mathematical model. Some previously obtained  results on  model's validation and bifurcation scenario are presented as well.  In section \ref{skur:secTwo} the  construction of impact trains and their statistical analysis are described. The final section contains the concluding remarks.

\section{ Experimental stand and its mathematical description}\label{skur:secOne}

\begin{figure}[t]
\begin{center}
\includegraphics[width=8 cm, height=4 cm]{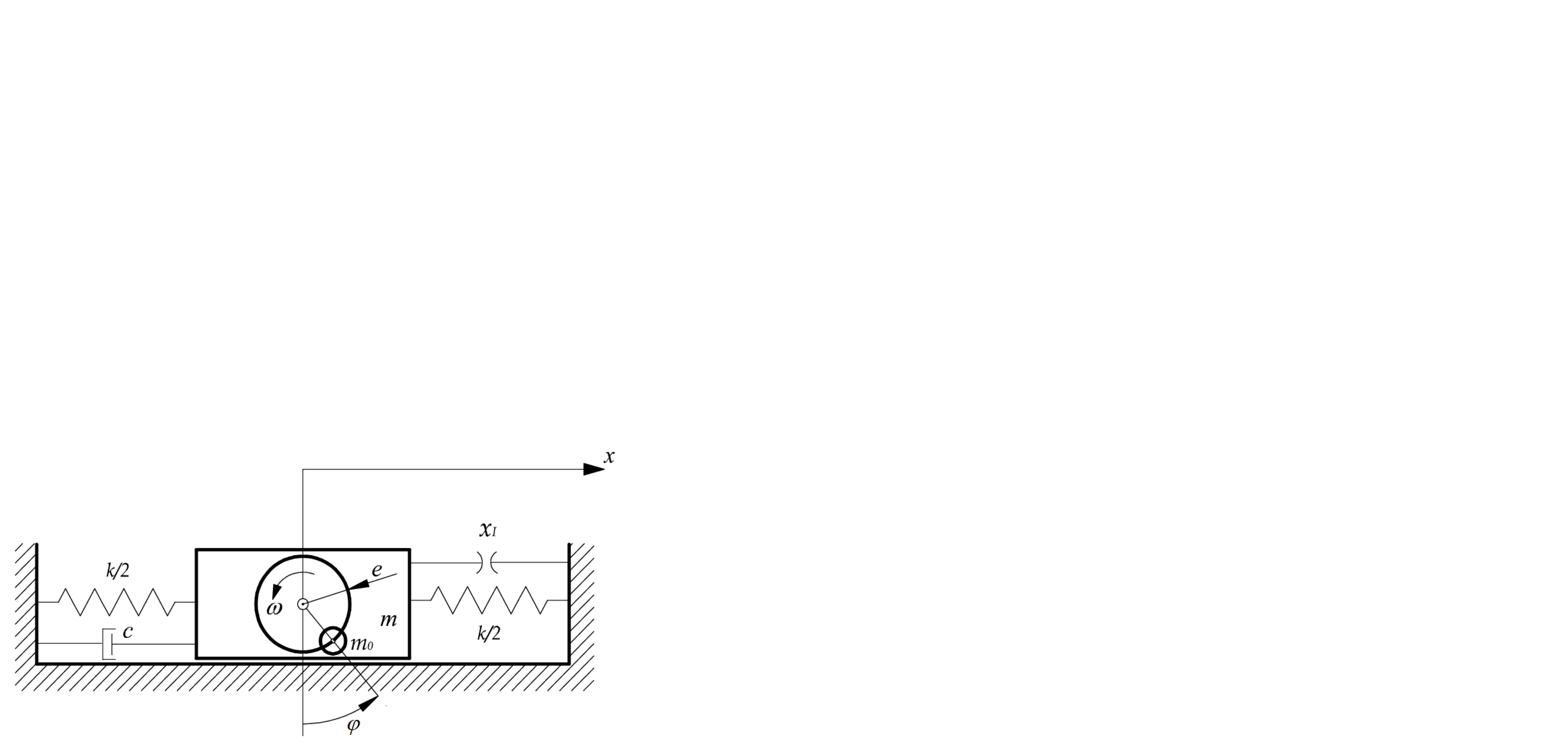}\hspace{0.5cm}
\end{center}
\caption{The schematic representation of the experimental stand \cite{Kudra2018}. 
} \label{skur:fig1}
\end{figure}

The base of our studies is the dynamical regimes observed at the experimental rig comprising an impacting oscillator. Since the detailed discussion  of this rig can be found in the paper \cite{Kudra2018},
we give the short description only. 
The experimental stand corresponding to the physical model presented in Fig.~\ref{skur:fig1} consists of a cart moving along a guide, integrated with linear ball bearing and Hall effect sensors. The cart is mounted elastically with the use of springs, while external forcing is generated by rotating unbalance mounted on the stepper motor with encoder situated on the cart. The position of the moving body is limited by an obstacle placed on the support. 
The bumpers are made of steel and have locally, near the impact point, forms of balls of radii equal to 25 mm. Moreover the centers of curvatures of both the bumpers and the center of the contact point lie on one line parallel to the direction of motion of the cart.
The experimental data is collected and processed by the use of National Instrument equipment and software.
To describe the behavior of the mechanical system from Figure \ref{skur:fig1}, the mathematical model 
was elaborated. 
It is assumed that the moving body possesses the total mass $m$ and its position is described by coordinate $x$. It is connected with the support by the use of linear spring of total stiffness coefficient $k$. The position $x=0$ corresponds to the resultant force in the springs equal to zero and a gap $x_I$ between the two bumpers on the right side of the cart. Angular position of the disk mounted on the cart is equal to $\varphi$, while its angular velocity $\omega=\dot \varphi$ is assumed to be constant or varying very slowly. The unbalance $m_0$ is placed on the radius $e$ of the disk.
 
  Concerning description of the impact, it is worth to note that one can encounter models assuming hard impacts, i.e. instantaneous events and often based on the coefficient of restitution, or the so called ''soft'' collisions. In the last case it should be distinguished between mechanical systems with soft obstacles and systems with collisions between hard elements modeled as locally deformable usually according to the Hertz model. The last case is used 
  in some resent works \cite{Okolewski2014,Okolewski2017} and 
  during our studies \cite{Kudra2018}.

To describe  the contact force during the impact with a compliant obstacle, a combination of spring and damper element (Kelvin-Voight viscoelastic model)
 is utilized. 
 In order to remedy  linear model's drawback  related to 
 the jump of the impact force at the beginning and end of the contact, Hunt and Crossley \cite{Hunt75} proposed nonlinear generalization of the linear model in the form
 $F=kh^
{n_1} +b h^{
n_2}  \dot h^{n_3}
$,
where $n_1$, $n_2$, $n_3$ are some parameters, $k$ is the  stiffness coefficient, $b $ is the  damping coefficient,  and $h$ stands for the relative penetration depth.

The total component of resistance proportional to velocity in both the spring and linear bearing is modeled in the form of viscous damper of coefficient $c$.
\begin{figure}[tbh]
\begin{center}
\includegraphics[width=6.5 cm, height=4.5 cm]{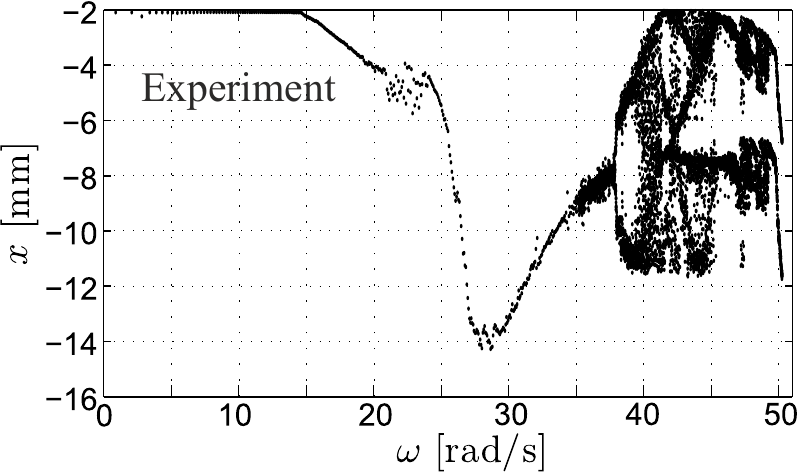}\hspace{0.5cm}
\includegraphics[width=6.5 cm, height=4.5 cm]{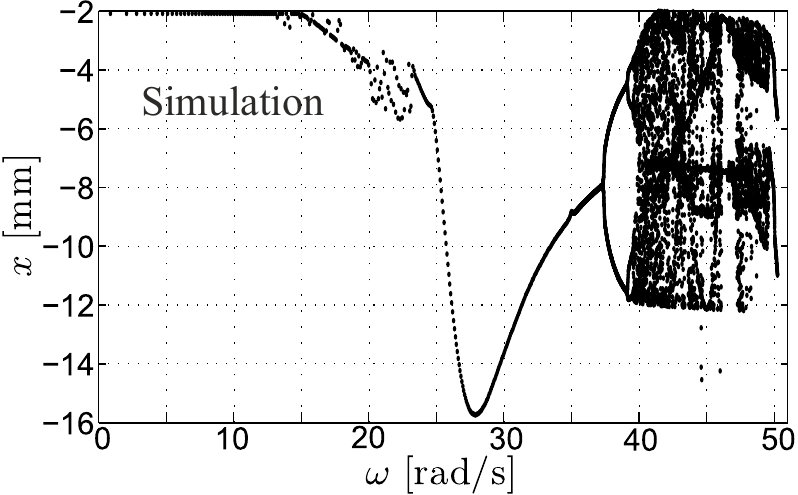}
\end{center}
\caption{The bifurcation diagrams constructed  experimentally (left panel) and numerically (right panel) at increasing $\omega$ and  obstacle position $x_I=-2.086$~mm \cite{Kudra2018}. 
} \label{skur:fig2}
\end{figure}
It is also
assumed that there exists some dry friction-like component of resistance between the cart and the guide, inside
the linear bearing, but it does not depend on normal
load.
The governing equations of the physical model take the following form
\begin{equation}\label{skur:model}
m\ddot x +kx +F_R (\dot x) + F_I (x, \dot x) =m_0 e \omega^2 \sin \varphi,
\end{equation}
where  the resistance force reads
\begin{equation*}
F_R=c\dot x+T \frac{\dot x}{\sqrt{\dot x^2+\varepsilon^2}},
\end{equation*}
while the impact force is
$F_I=
       k_I ((x-x_I)^{n_1}+b(x-x_I)^{n_2} \mbox{sgn} \dot x |\dot x| ^{n_3})$ 
       for $ x-x_I \ge 0 $ and $ 
        (x-x_I)^{n_1}+b(x-x_I)^{n_2} \mbox{sgn} \dot x |\dot x| ^{n_3}\ge 0$, and $F_I=0$
otherwise.
In the work \cite{Kudra2018} the following set of parameters was estimated or assumed a priori leading to good agreement between numerical simulations and experimental investigations: $m=8.735$ kg, $m_0 e=0.01805$ kg$\cdot$m, $k=1418.9$ N/m, $c=6.6511$ N$\cdot$s/m, $T=0.63133$ N, $\varepsilon =10^{-6}$ m/s, $k_I=2.3983\cdot  10^8$ N/ m$^{3/2}$, $b=0.8485$ m$ ^{-n_3 }$ s$^{n_3}$, $n_1=n_2=3/2$, $n_3=0.18667$. In the paper \cite{Kudra2018}, the model (\ref{skur:model}) specified by this set of parameters is called the model BC and provides  the optimal one from a group of models tested in the work \cite{Kudra2018}.

 The comparison of experimental and numerical studies is presented in Fig.~\ref{skur:fig2}, which exhibits two bifurcation diagrams obtained numerically and experimentally for the obstacle position $x_I=-2.086\cdot 10^{-3}$ m, based on Poincare maps defined by sections $\varphi =2\pi i$, $i=1,2,3\dots $.  The analysis carried out  in the present work is used the parameters fixed above and  position of the limiter $x_I=-2.086\cdot 10^{-3}$ m.

Here it should be mentioned that the paper \cite{Kudra2018} contains the description for experimental investigations of the oscillating system with impacts, construction and validation of different mathematical models, and bifurcation diagrams exhibiting the general properties of observed oscillating regimes.

In the presented research, we study  the mathematical model which is the best fitting the experiment. Moreover, we develop the tools for examining the attractors (especially chaotic) in details, namely their statistic properties, using the impact sequences. It turns out, these sequences are extremely informative. In combination with their simple extraction from the physical model, the impact sequences can have great potential for future applications.

\section{Studies of attractors' properties}\label{skur:secTwo}

\begin{figure}
\begin{center}
\includegraphics[width=6.5 cm, height=4.5 cm]{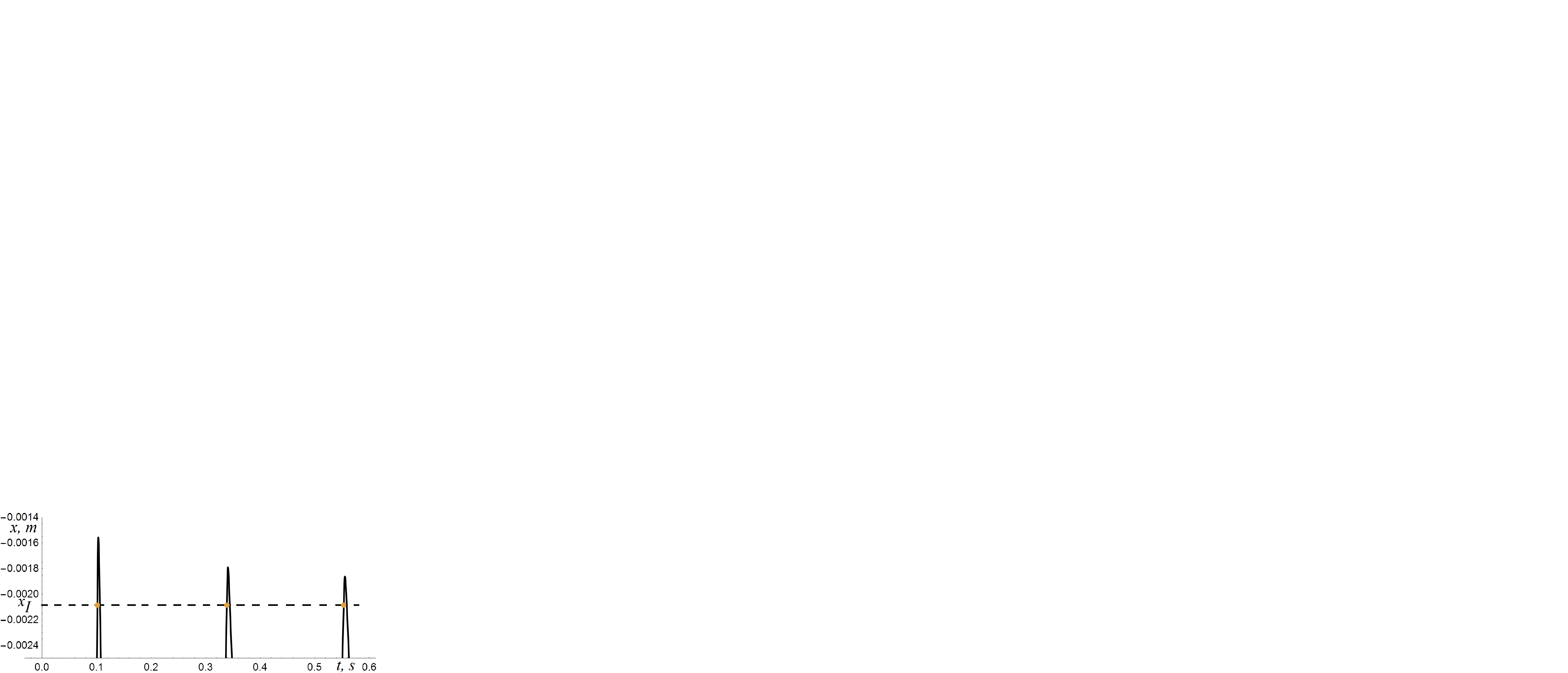}\hspace{0.5cm}
\includegraphics[width=6.5 cm, height=4.5 cm]{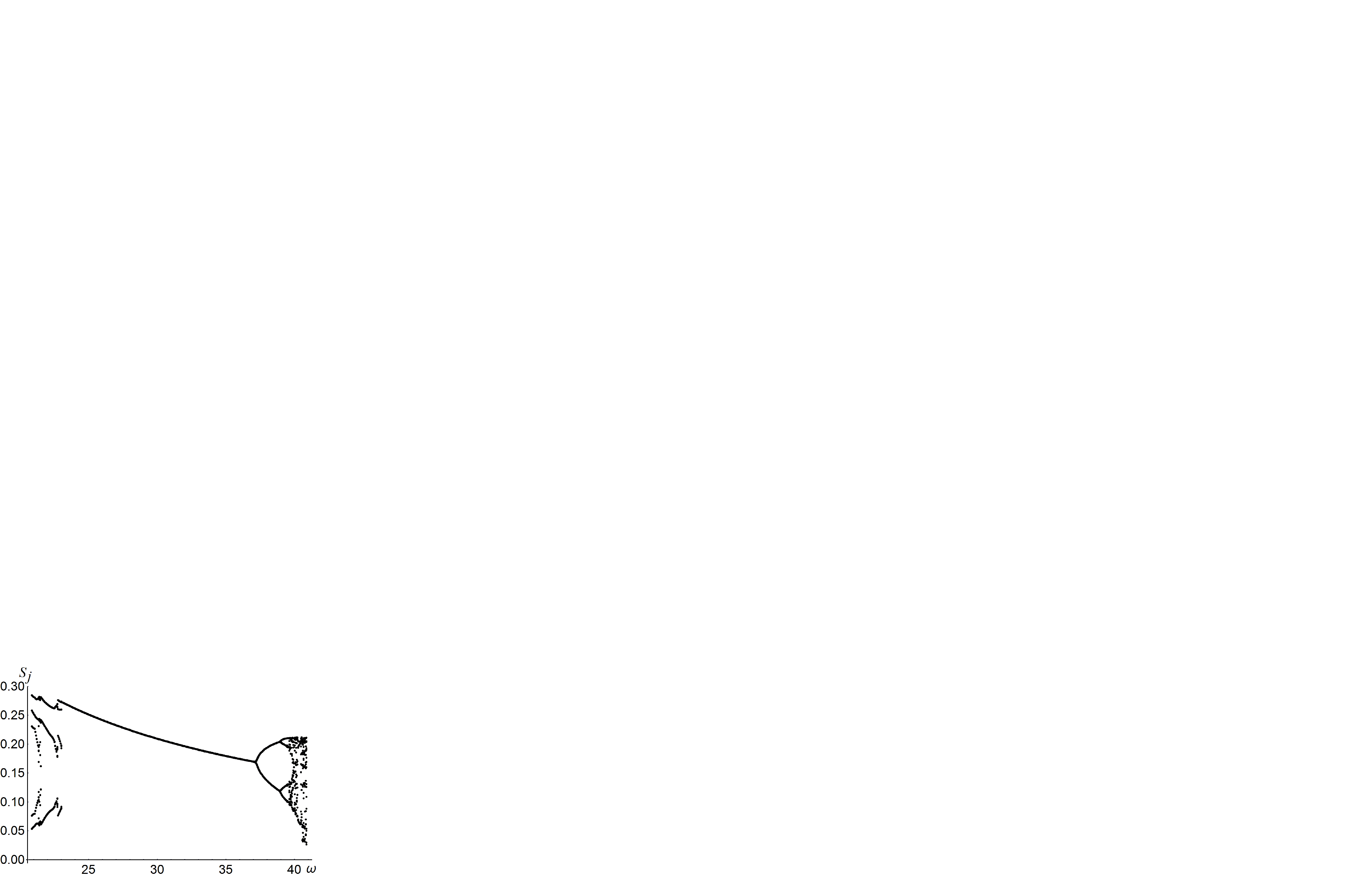}\\
(a)\hspace{7 cm} (b)
\end{center}
\caption{The construction of impact map and inter-impact intervals  (a), and bifurcation diagram  for  inter-impact intervals (b) at decreasing $\omega$. 
} \label{skur:fig3}
\end{figure}

  Now we are interested in the properties of attractors and their 
  bifurcations via sequences of points generated  by the stops' impacts. These sequences form the impact map and can be regarded as a sort of Poincare map. To construct this map, we analyze the profile of $x$ component observing  the intersections of increasing trajectory with the level $x=x_I$ (Fig.\ref{skur:fig3}a). These time moments of  impacts  define  the beginning of collision and can be  easily   derive numerically. 

As a result, the impact map as the set of points $\{T_j: x(T_j)=x_I\}$ can be extracted from the temporal profile of $x$ during the numerical integration of model (\ref{skur:model}). It is obvious that periodic regimes produce the periodic sequence $\{T_j\}$, whereas chaotic attractors generate the stochastic sequences. These points define the inter-impact  intervals $S_j=T_{j+1}-T_j$.

Using the impact map, the rearrangement of the phase space of equation (\ref{skur:model}), when model parameters vary, can be studied by means of  bifurcation diagram. To make it,  we put the control parameter $\omega$ along horizontal axis, whereas the values of interval's width $S_j$ are putted along vertical one.  The resulting diagram obtained at the decreasing frequency $\omega$  from $40.86$~rad/s  is plotted in Fig.~\ref{skur:fig3}b. 

\begin{figure}[tbh]
\begin{center}
\includegraphics[width=6.5 cm, height=3.5 cm]{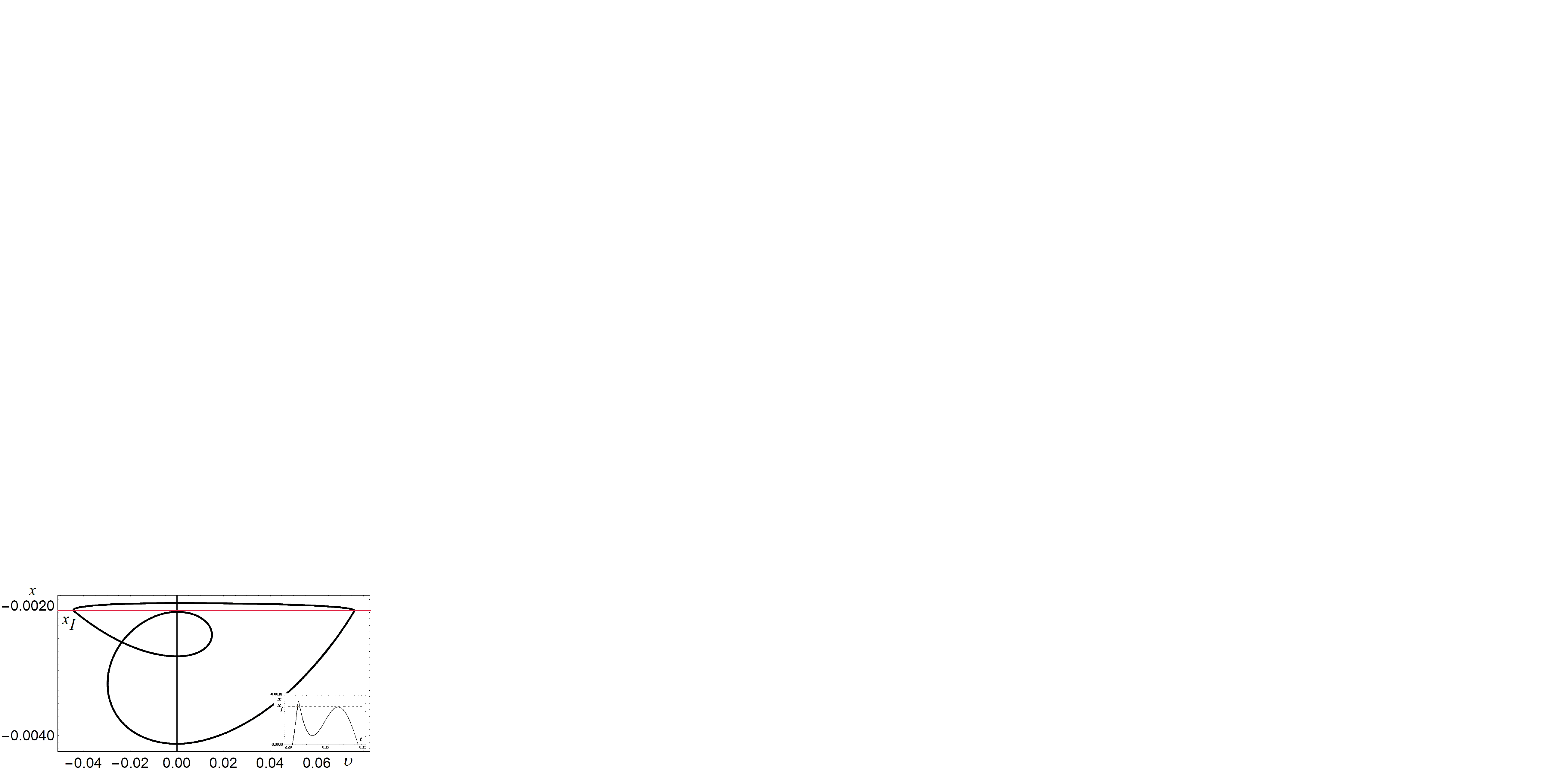}\hspace{0.5cm}
\includegraphics[width=6.5 cm, height=3.5 cm]{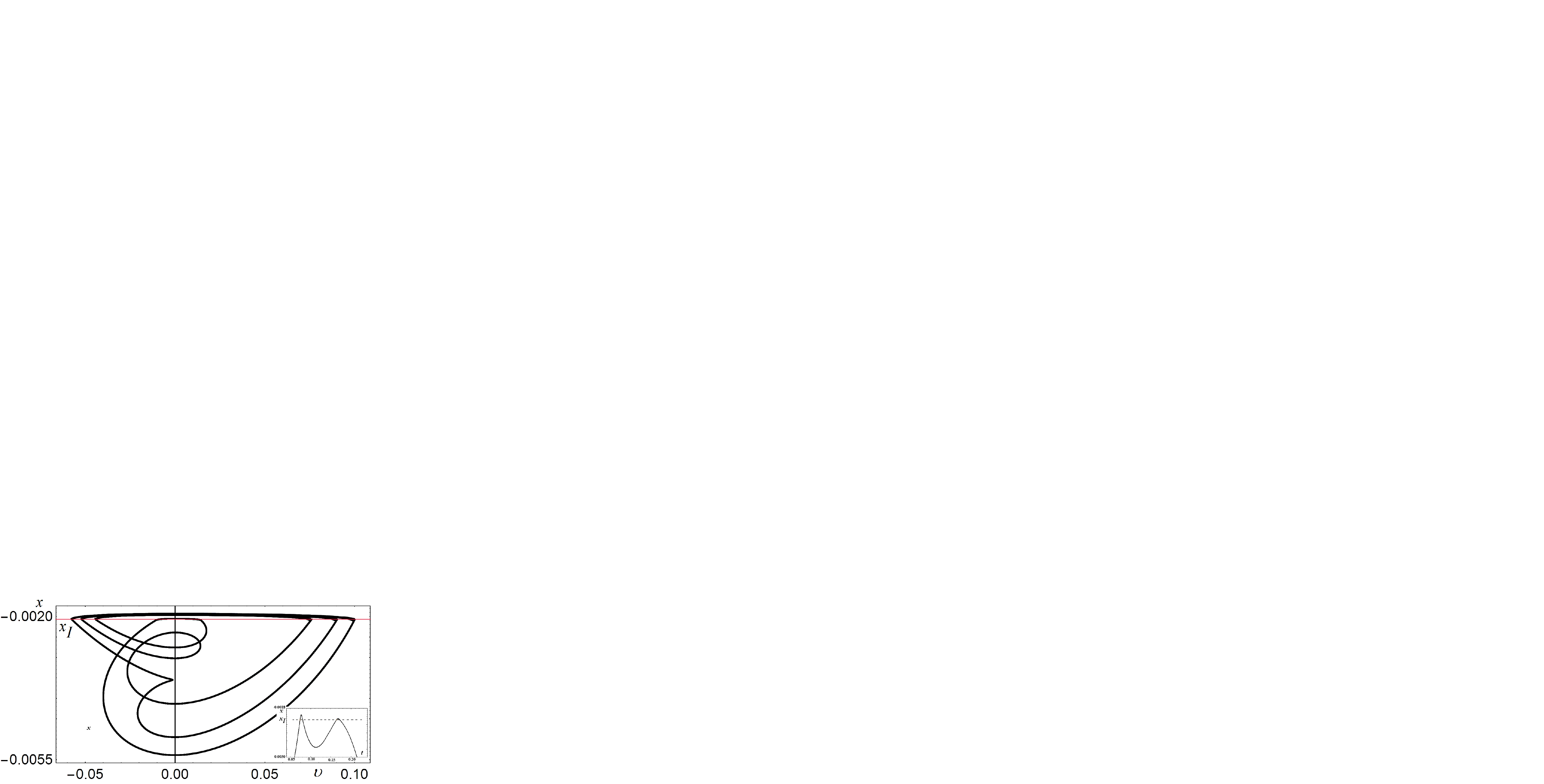}\\
(a)\hspace{7 cm} (b)\\
\includegraphics[width=6.5 cm, height=3.5 cm]{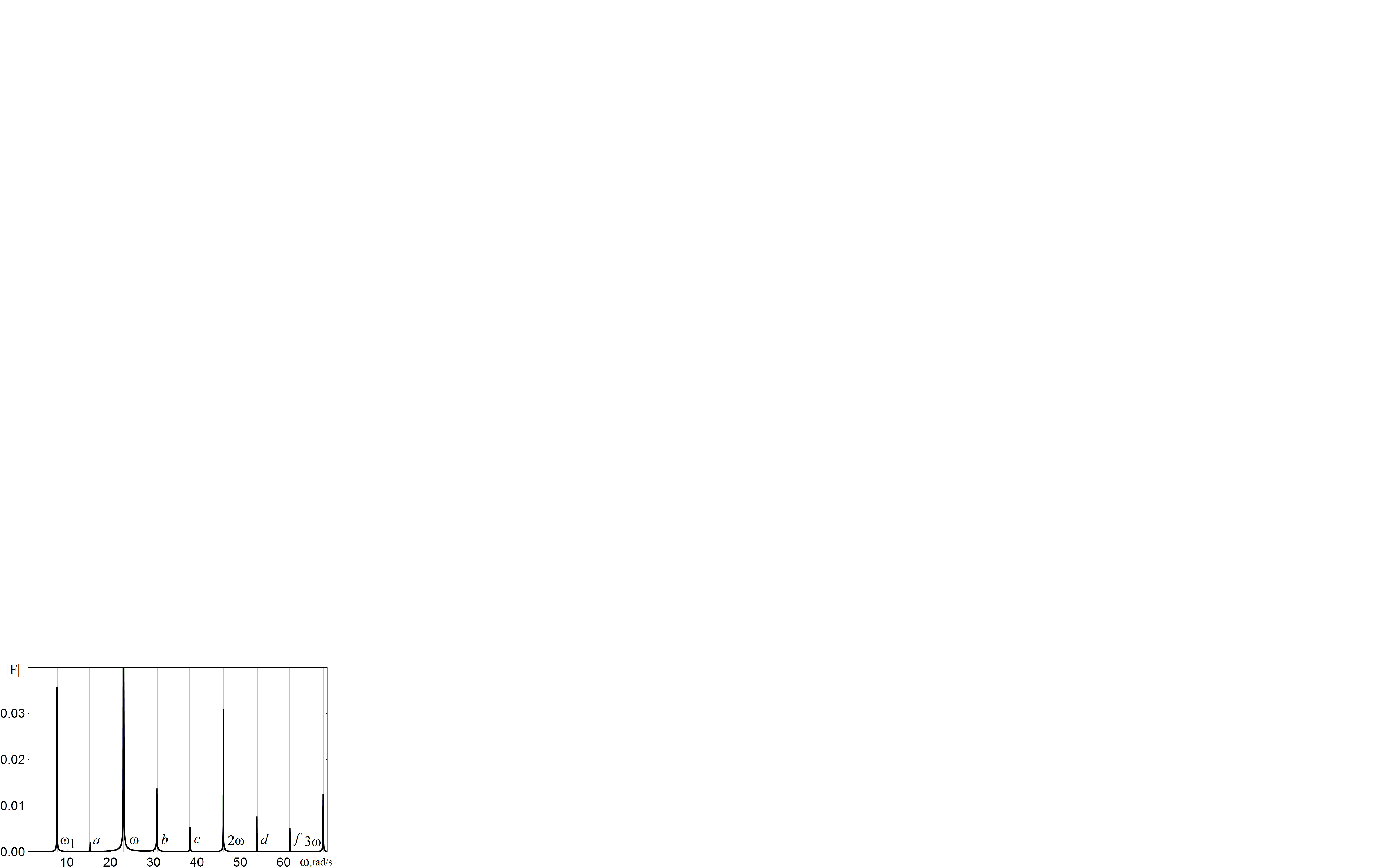}\\
  (c)
\end{center}
\caption{Periodic trajectory at $\omega=23.05$~rad/s just before the grazing bifurcation (a) and  after the bifurcation at $\omega=23.03$~rad/s  (b). The insets exibit the parts of profiles where the grazing occurs. The Fourier spectrum (c) for the attractor depicted in panel (b).  The symbols $a$, $b$, $c$, $d$, $f$ designate the frequency combinations $\omega \pm \omega_1$, $2\omega\pm\omega_1$, and $3\omega-\omega_1$.
} \label{skur:fig4}
\end{figure}

From this diagram it follows that reversed period doubling bifurcations occur at frequencies closing to starting $\omega=40.86$~rad/s.  The left part of the diagram exhibits specific type of bifurcations inherent basically in the models with impacts. In particular, at about $\omega=23$~rad/s     the additional branches of bifurcation curve appear beneath the main curve corresponding to the limit cycle  existence. In this case we deal with the grazing bifurcation \cite{Grazing1994} when the trajectory touches the point of impact  with zero velocity as it is shown in the inset of Figs.~\ref{skur:fig4}. The phase portraits in Figs.~\ref{skur:fig4} correspond to attractors just before the bifurcation at the external frequency  $\omega=23.05$~rad/s (Fig.~\ref{skur:fig4}a) and after it at $\omega=23.03$~rad/s (Fig.~\ref{skur:fig4}b). Note that according to Fourier spectral analysis, 
the former regime possesses the modes $\omega \cdot k$, $k=1,2,\dots$ and latter regime, in addition to $\omega \cdot k$, is endowed by the $\omega_1<\omega$ and combinational frequencies $\omega \cdot k \pm \omega_1$. We thus run into the regime described by the quasiperiodic function nonlinearly depended on two periodic functions of periods $2\pi/\omega$ and  $2\pi/\omega_1$ \cite{BergeQuasi}. 

\subsection{Periodic regime} 
Now consider the properties of periodic attractor existed at $\omega=23.03$~rad/s (Fig.~\ref{skur:fig4}b) using the impact map. Let us integrate model (\ref{skur:model}) during 100~s and get the sequence $T_j$ containing 489 elements.    Note that  \emph{Mathematica}'s procedure for capturing the curves intersections requires 
specific value of \textsf{MaxStepSize}  option, i.e. \textsf{MaxStepSize} =0.01 for periodic sequence and \textsf{MaxStepSize} = $10^{-5}$ for chaotic impact trains in order to avoid the omissions of impacts. 

\begin{figure}[h]
\begin{center}
\includegraphics[width=6.5 cm, height=4.5 cm]{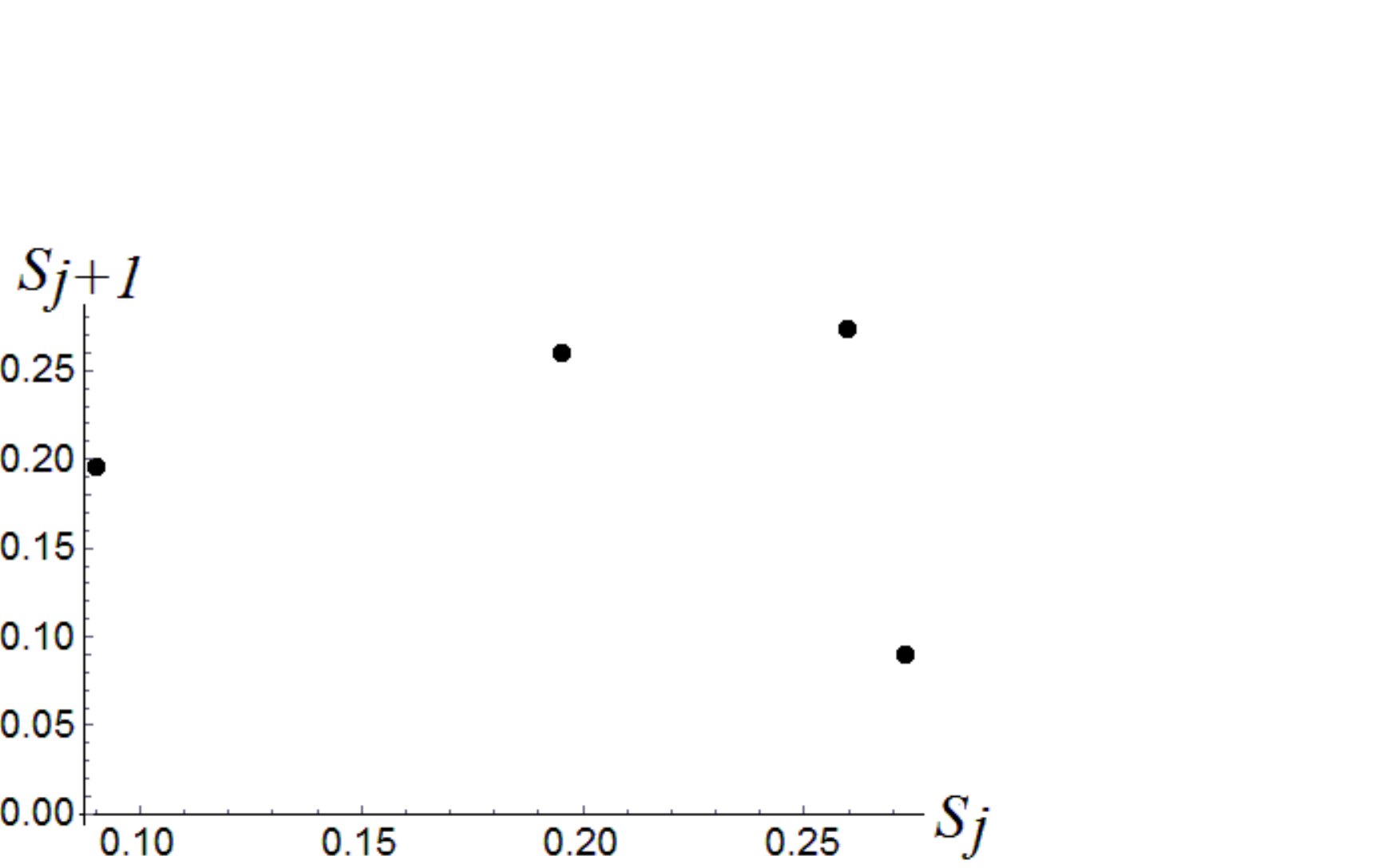}\hspace{0.5cm}
\includegraphics[width=6.5 cm, height=4.5 cm]{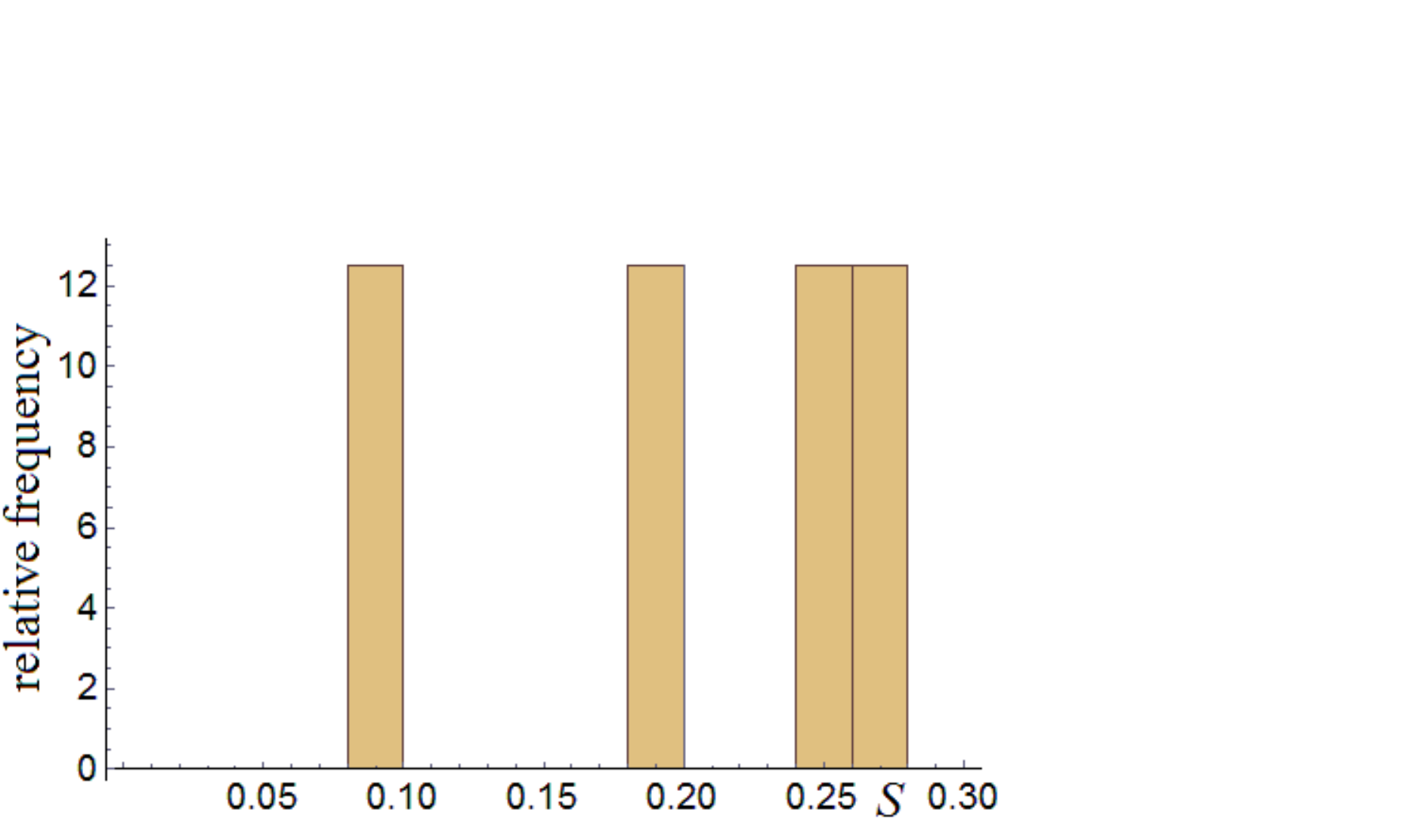}\\
(a)\hspace{7 cm} (b)
\end{center}
\caption{Successive iterations (a) of Poincare return map for the periodic trajectory at $\omega=23.05$~rad/s and corresponding histogram (b). 
} \label{skur:fig5}
\end{figure}

Constructing the sequence of impact intervals $\{S_j\}$ via aforementioned method, the  successive iterations $S_{j+1}=f(S_j)$ can be arranged (Fig.~\ref{skur:fig5}a).

We see that  only four different points are distinguished. This means that only four distinct intervals (or temporal scales) are present in the interval sequence. The histogram built for these points shows that the number of  $S_j$ of each width  is equal.   It is worth to note that, together with the analysis of Fourier spectrum and classic Poincare section,  impact map allows one to supplement  information on the trajectory behavior. In particular,  it is obvious that the sum of arbitrary four successive elements of $S_j$ equals the period of trajectory. In other word,  the sum of four coordinates of histogram's bins gives 0.8184. 

\subsection{ Chaotic regime}

\begin{figure}
\begin{center}
\includegraphics[width=6.5 cm, height=4.5 cm]{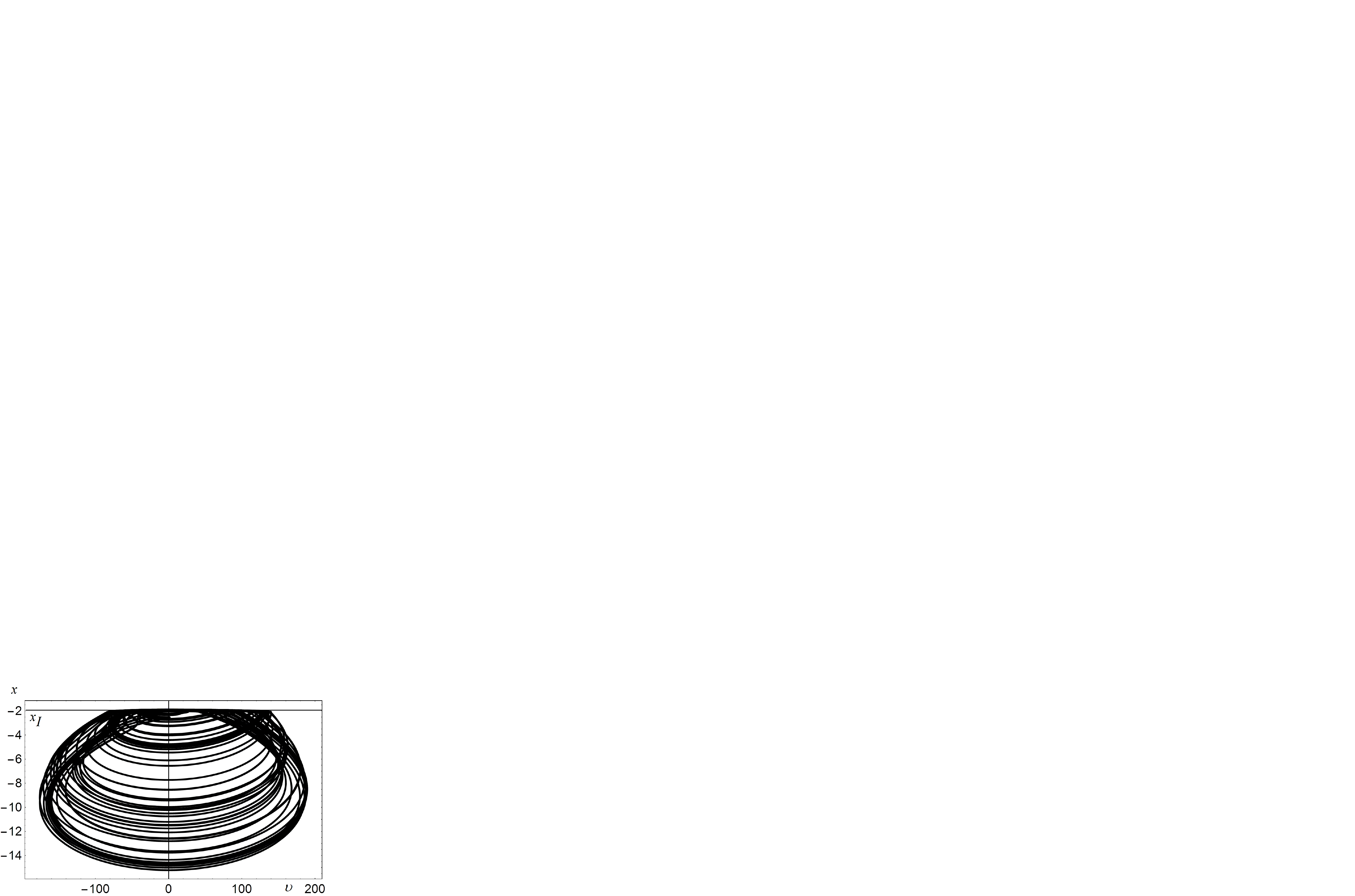}\hspace{0.5cm}
\includegraphics[width=6.5 cm, height=4.5 cm]{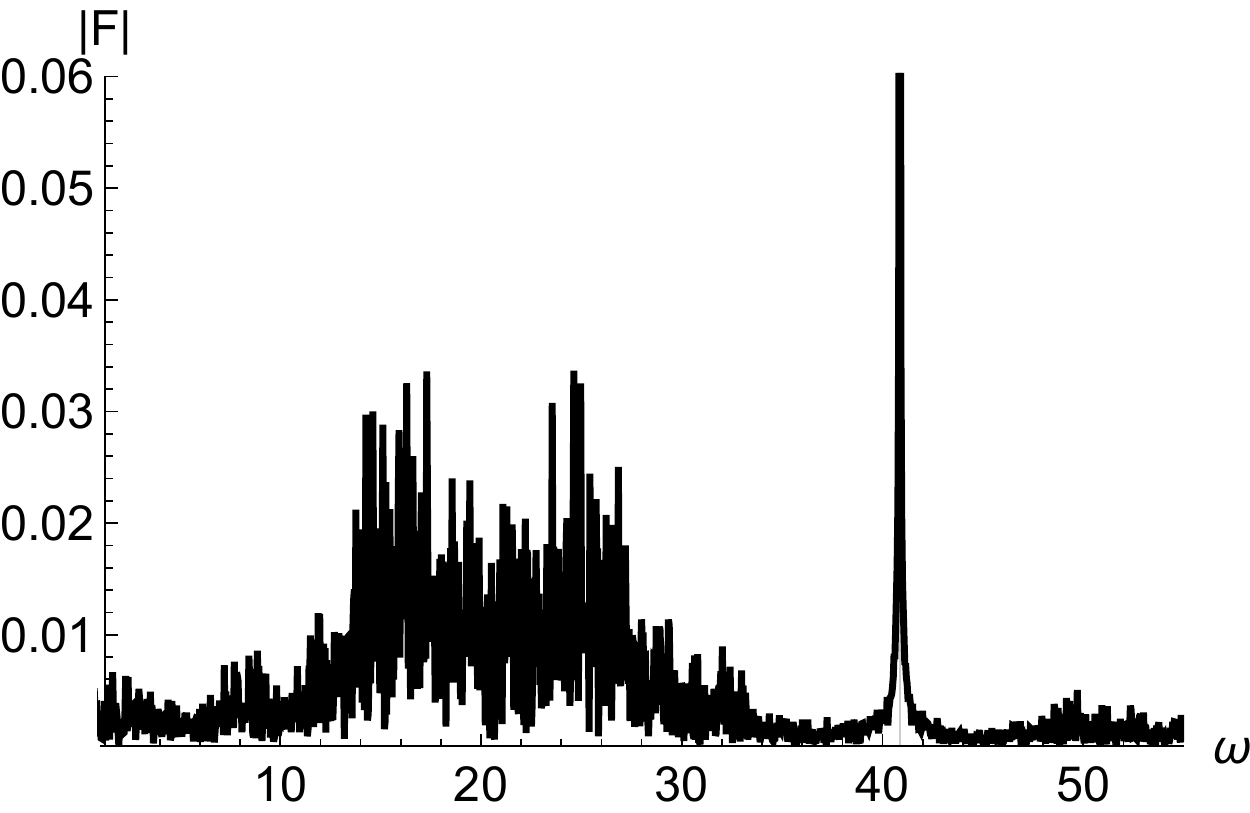}\\
(a)\hspace{7 cm} (b)
\end{center}
\caption{The phase portrait (a) of the chaotic attractor and corresponding Fourier spectrum (b)  of $x$-component. 
} \label{skur:fig6}
\end{figure}

\begin{figure}
\begin{center}
\includegraphics[width=6.5 cm, height=4.5 cm]{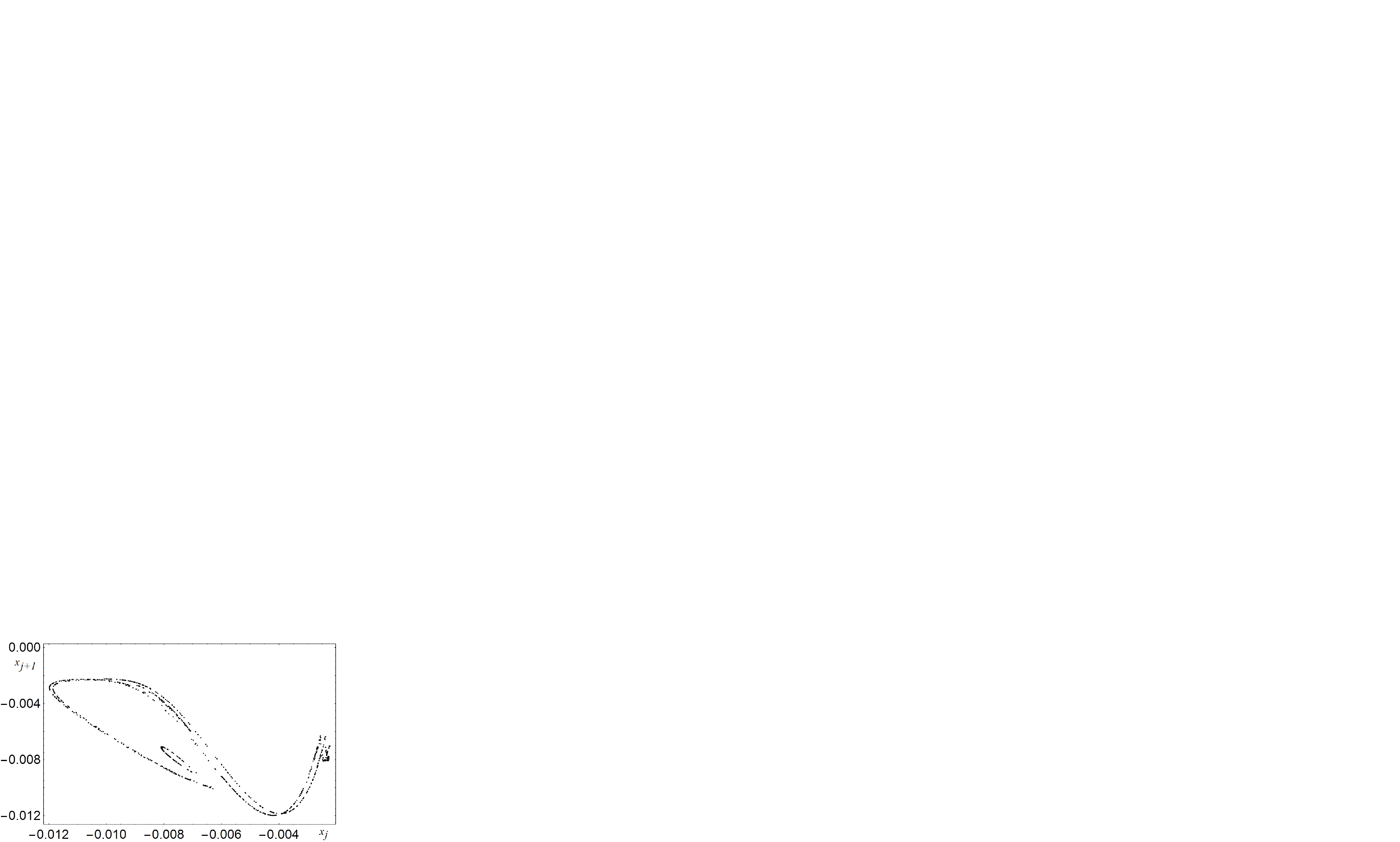}\hspace{0.5cm}
\includegraphics[width=6.5 cm, height=4.5 cm]{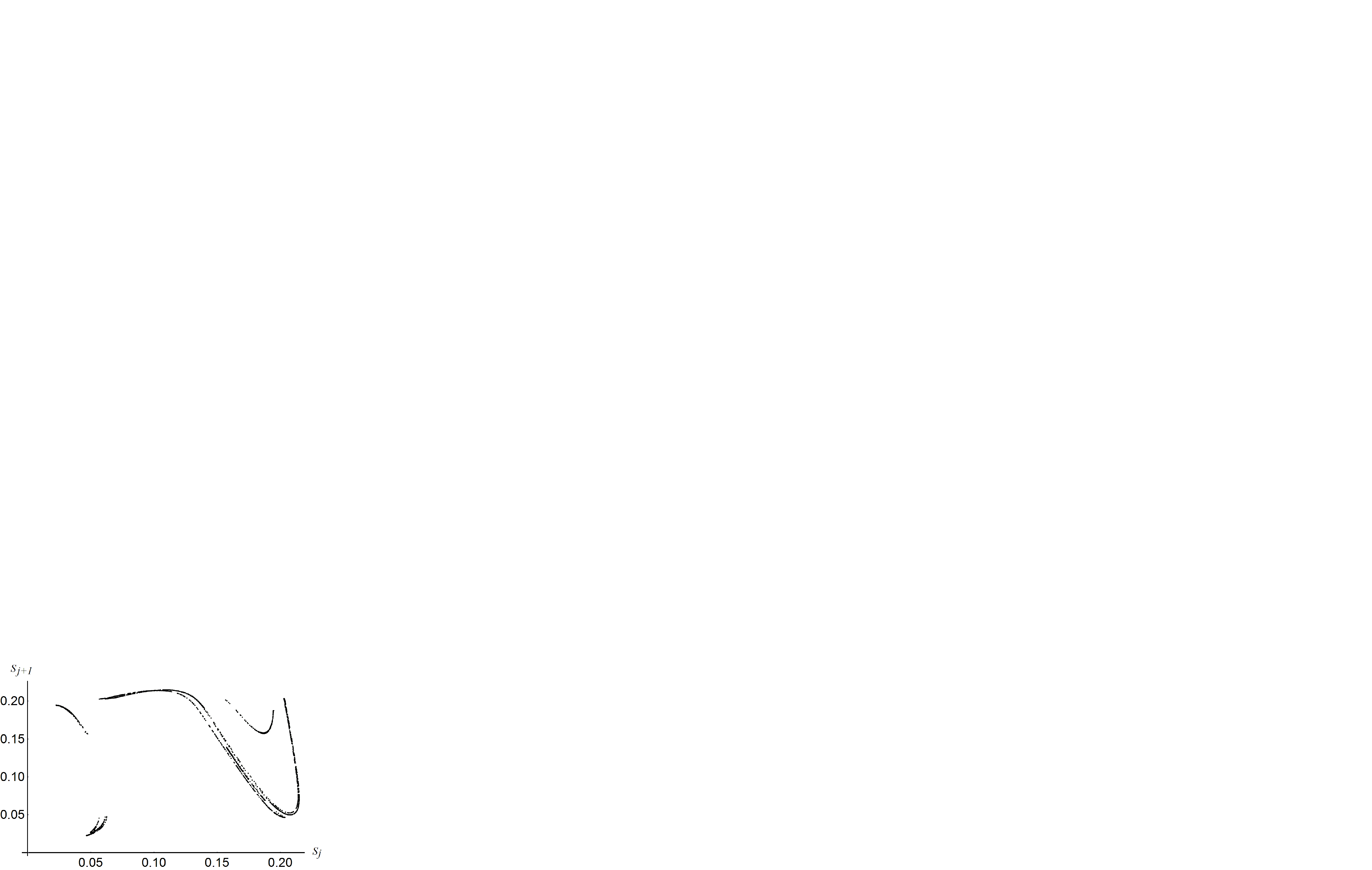}\\
(a)\hspace{7 cm} (b)
\end{center}
\caption{Successive iterations of the Poincare return map (a) and impact map (b).
} \label{skur:fig7}
\end{figure}

Consider the properties of impact map in a chaotic regime when statistical features of the regime manifest brightly.  For studies let us chose the  case at $\omega=40.86$~rad/s (period 0.154). Recall \cite{Kudra2018} that the trajectory after some transition time leads to the chaotic attractor, the phase portrait of which is depicted in   Fig.~\ref{skur:fig6}a.  Corresponding Fourier spectrum for this chaotic trajectory (Fig.\ref{skur:fig6}b) possesses one substantial extremum corresponding to the  frequency $\omega$ of forcing, whereas at smaller  frequencies the spectrum contains  dense set of  excited frequencies. Also important characteristic of chaotic regime is the Poincare map coinciding with the set of points $x(t)$ in the section planes $\varphi=2\pi i$, $i=1,2,\dots$. Grouping them into the sequence $\{x_j;x_{j+1}\}$, we obtain Fig.~\ref{skur:fig7}a \cite{Kudra2018}.

The similar sequence can be constructed for impact map after rearrangement of  the sequence $S_i$ into the set $\{S_j; S_{j+1}\}$ (Fig.~\ref{skur:fig7}b). Both diagrams have fractal nature but the successive iterations for impact map form the separated branches of map. Note that similar shape of map graph  is encountered during the studying the  return time map in the Rossler system \cite{AnishchenkoPRE2001} and pendulum system from \cite{Slade97}.

In this case, unlike the periodic regime,  the 
sequence of points generated during the impacts is  stochastic. Therefore, the statistic properties of these impacts train $T_j$ are in question. Considering the number $N(t)$ of impacts in the time interval $(0,t)$, we lead to the studies of distribution of $N$ under the auxiliary assumptions: stationarity, independent increments and orderlines \cite{Probability2000}. 

At first, let us check the stationarity of $T_j$. To do this, the number of impacts should be assessed at different intervals of time. We take total time interval $t=400$~s and divide it in equal five intervals. Counting the number of impacts in each time interval, one can obtain  
 $\{572 , 571 , 579 , 579 , 574\}$ impacts (total number of impacts is 2875). Since the number of impacts in each time interval is almost equal then we can state that the process is 
 \emph {stationary}. 
 
 We also suppose that the number $N$ depends on the length of time interval only and impacts do not appear in groups providing the implementation of another assumptions concerning stochastic process. 
 
As it is well known, the random variable $N$ obeying aforementioned properties is described by the Poisson distribution $P(N(t)=n)=(\lambda t)^n \exp(-\lambda t)/n!$, where $\lambda$ is the rate of Poisson process \cite{Probability2000}. Note that many natural temporal processes obey this law, for instance, the number of earthquakes, $\beta$-particles after radioactive decay, spikes in neuron activity.  To define the type of distribution, it is useful to investigate the coefficient of variation $C_V$ \cite{Gerstner2014}
\begin{equation}
C_V^2=\frac{D[N]}{M[N]},
\end{equation}     
   where $D[N]$ and $M[N]$ are the variance and expected value of variable $N$.
   
\begin{figure}
\begin{center}
\includegraphics[width=6.5 cm, height=4.5 cm]{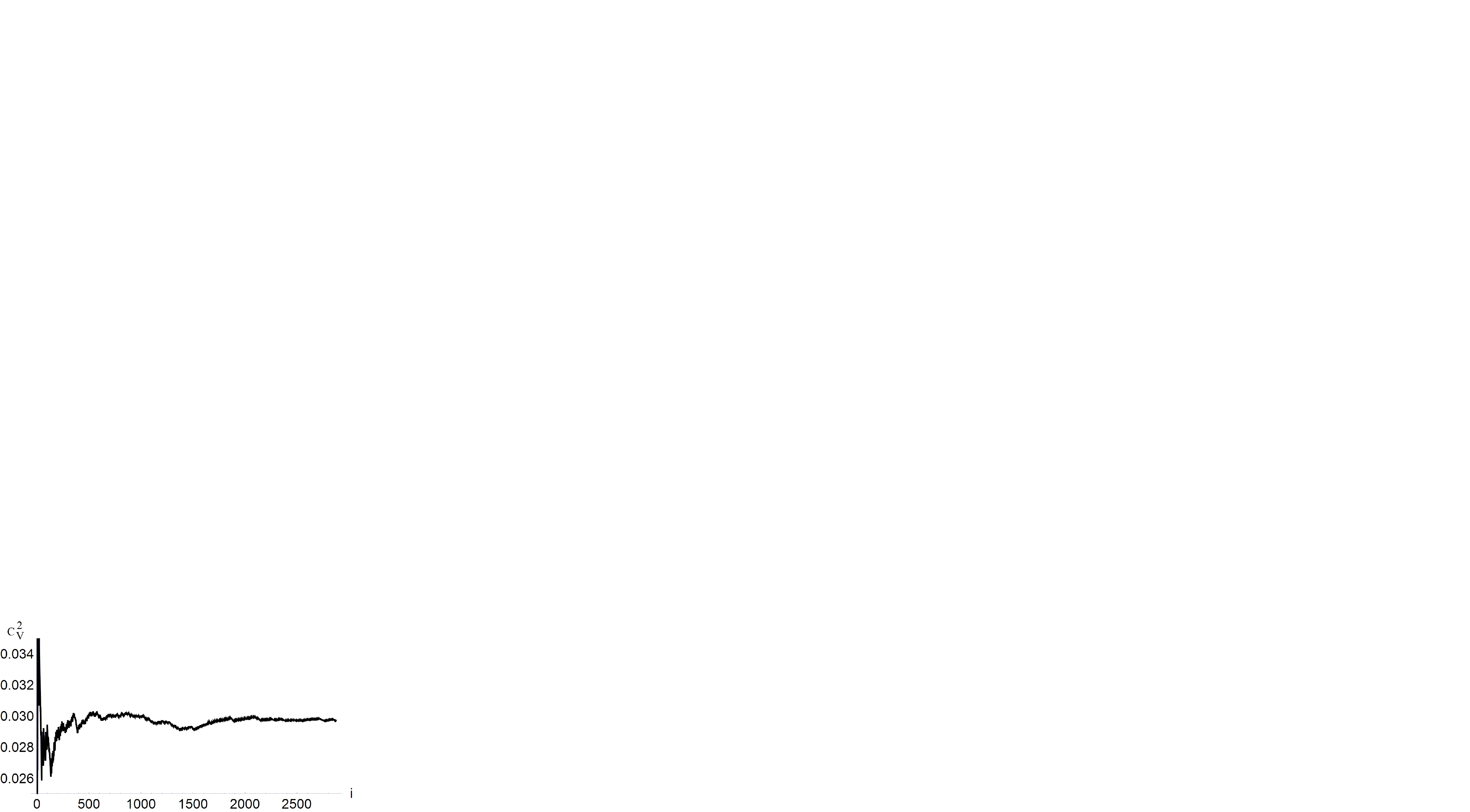}\hspace{0.1cm}
\includegraphics[width=6.5 cm, height=4.5 cm]{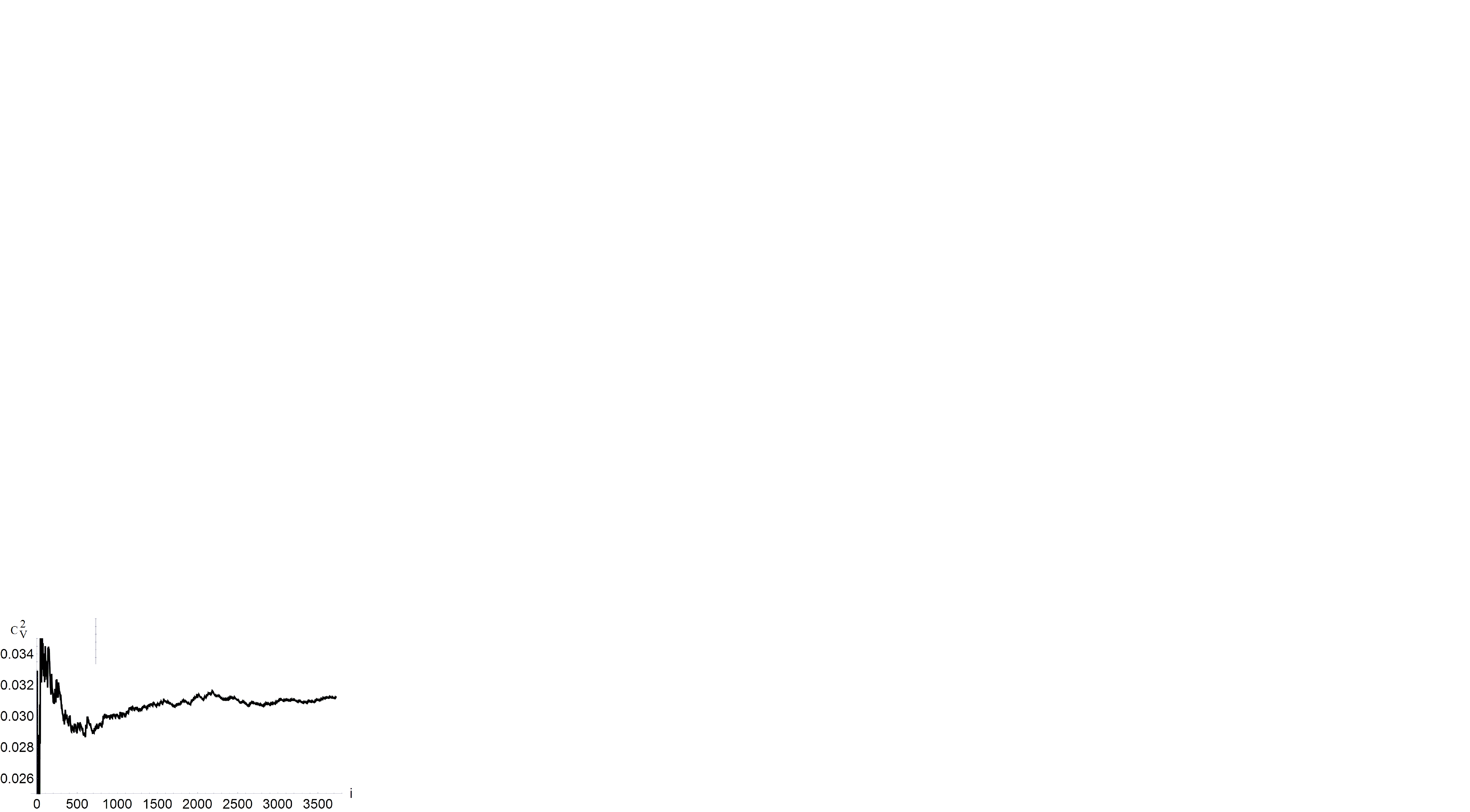}\\
(a) \hspace{5 cm} (b)
\end{center}
\caption{Coefficient of variation $C_V^2$ for numerical model (a) and experimental (b).
} \label{skur:fig8}
\end{figure}

   For the Poisson process $C_V^2=1$ \cite{Probability2000}. Estimation of $C_V^2$ with the help of sequence of inter-impact intervals for model (\ref{skur:model}) gives the Fig.~\ref{skur:fig8}a. From this it follows that for large sample the quantity $C_V^2$ tends to its limiting value about 0.03. Since $C_V^2(i>2000)<1$, then the impact sequence behaves itself more regular in comparison with the Poisson process \cite{Gerstner2014}.      
   
   The coefficient of variation $C_V^2$ also is derived for the experimental sample of ISI (Fig.~\ref{skur:fig8}b). The $C_V^2$ profile stabilizes for large $i$ in the vicinity of 0.03 that  a little bit higher than for the numerical ISI. Nevertheless, the general conclusion on the  behavior similarity of both ISI sequences is valid. 
   
\begin{figure}
\begin{center}
\includegraphics[width=6.5 cm, height=4.5 cm]{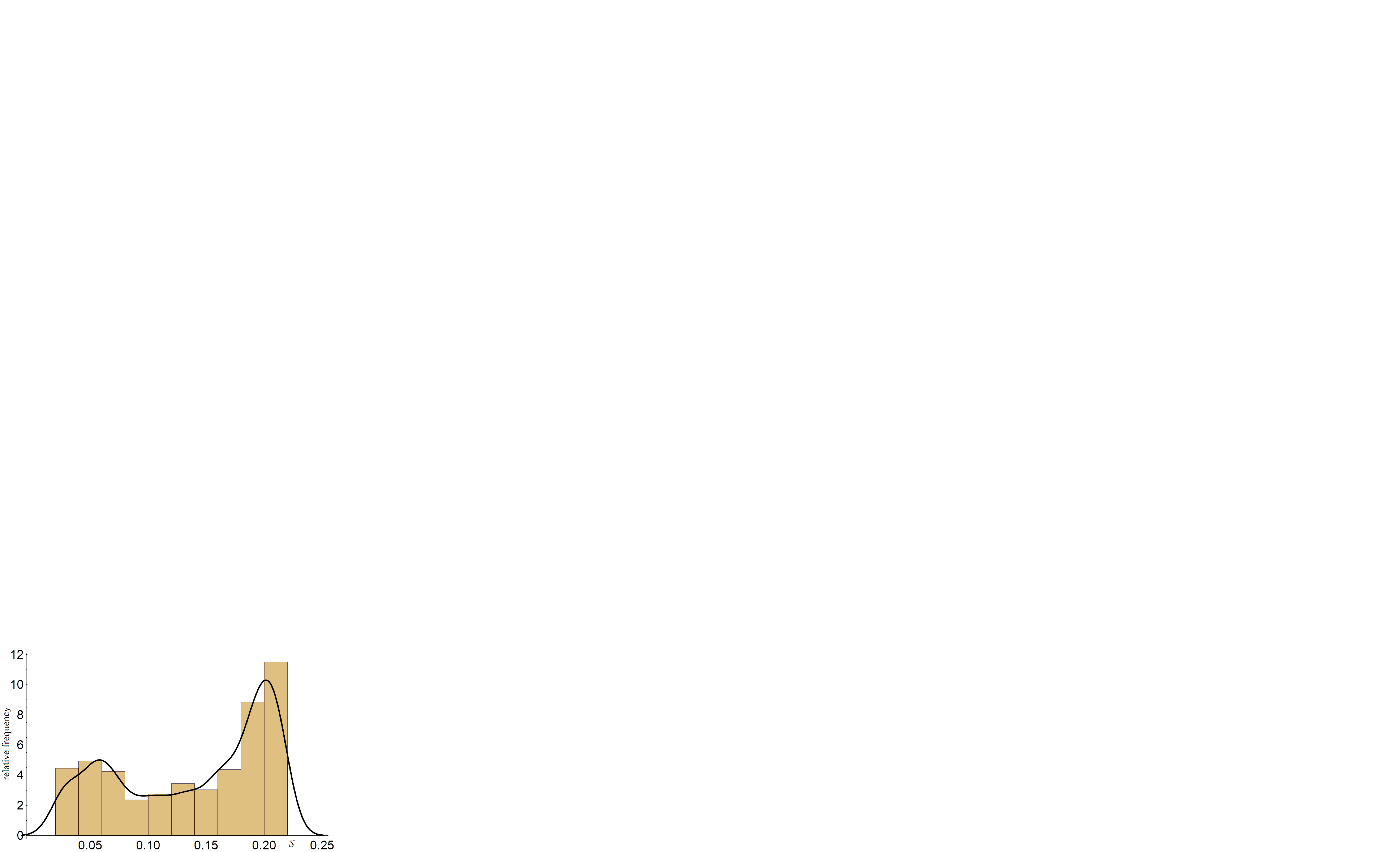}\hspace{0.5cm}
\includegraphics[width=6.5 cm, height=4.5 cm]{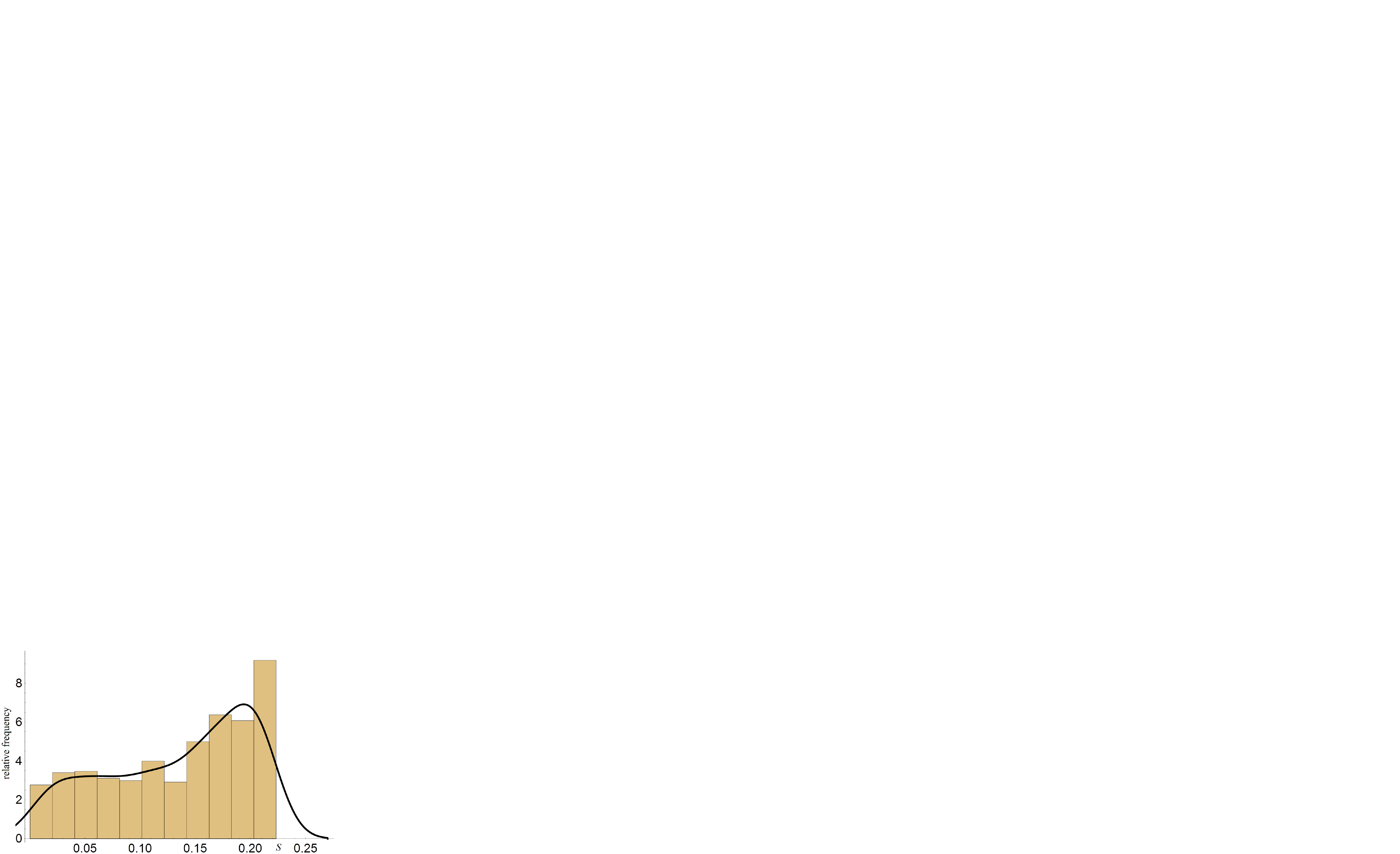}\\
(a)\hspace{7 cm} (b)
\end{center}
\caption{The relative frequency histogram   of inter-impact intervals and its smoothed version estimated  numerically (a) and experimentally (b). 
} \label{skur:fig9}
\end{figure}

Consider now the relative frequency histogram  for ISI providing the estimation of a probability density function  of a random variable. Using the proper tools of \emph{Mathematica} system, we arrange the 2875 intervals in the ten bins and obtain the histogram in Fig.~\ref{skur:fig9}a. The smooth line connecting  the histogram bars highlights clearly the two distinct maxima at the points $T_1=0.05$~s and $T_2=0.2$~s.  These quantities $T_{1,2}$ can relate to some temporal scales in the chaotic profile of $x$. Note that external forcing  is characterized with the period $T=2\pi/\omega=0.15$~s which does not coincide with $T_{1,2}$.  It is also worth to note that the histogram looks like a bimodal distribution which often encounters in nature \cite{AnishchenkoPRE2001,Levine77,Sauer95}. Till now, the problem of approximation for derived bimodal distribution was not considered, although some progress  in this regard has been  achieved  \cite{Levine77}.  We also construct the relative frequency histogram (Fig.\ref{skur:fig9}b) for the experimental data. This histogram possesses the substantial maximum for long intervals and weakly expressed extremum for short intervals. From this follows that numerical and experimental histograms have similar shapes. It should be noted that the experimental data contains noisy components. But, as it is shown in \cite{Sauer95},   noise incorporation in time series leads to the degenerate of bimodal distribution into unimodal.    
\begin{figure}[bth]
\begin{center}
\includegraphics[width=6.5 cm, height=4.5 cm]{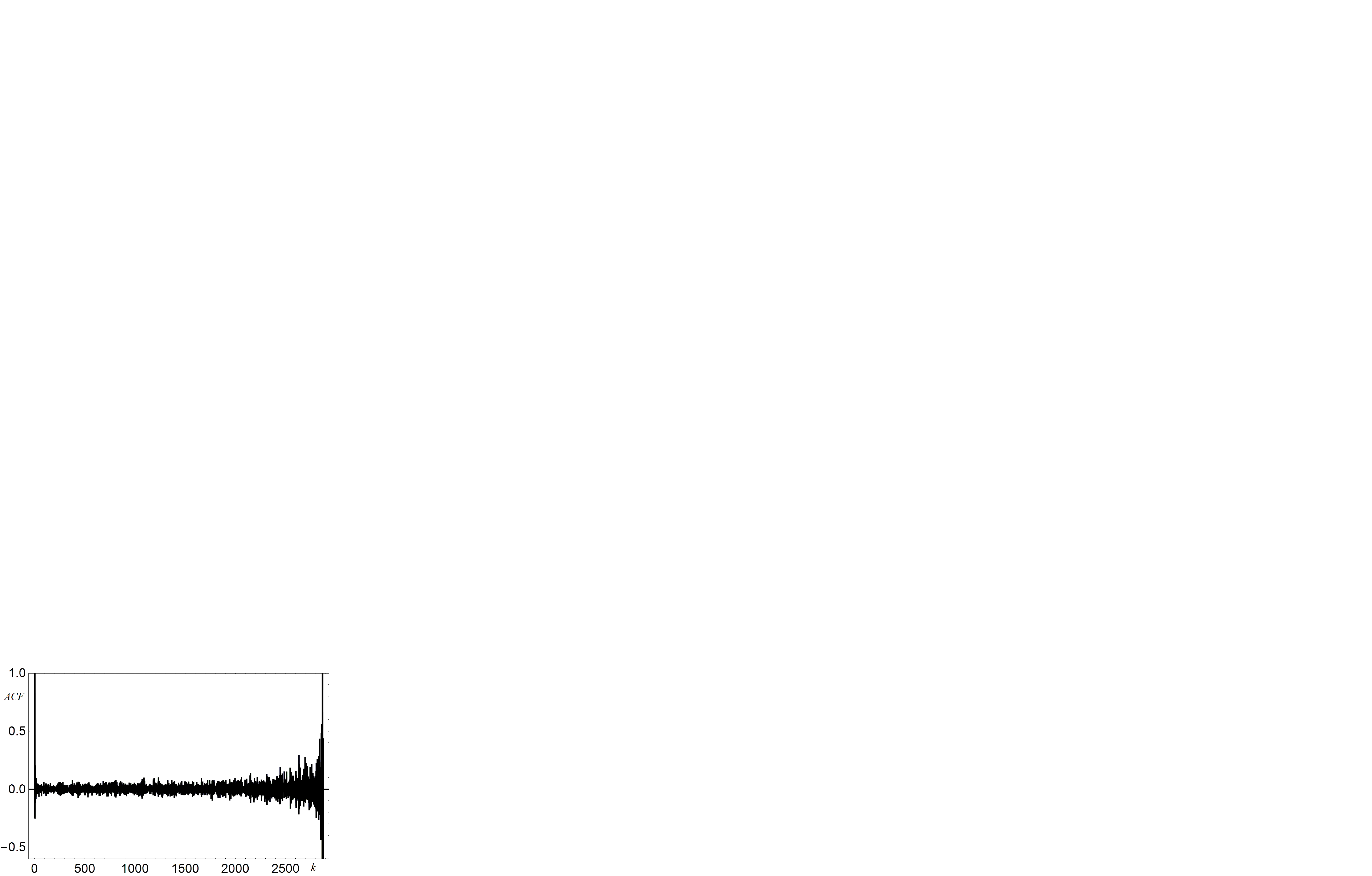}\hspace{0.5cm}
\includegraphics[width=6.5 cm, height=4.5 cm]{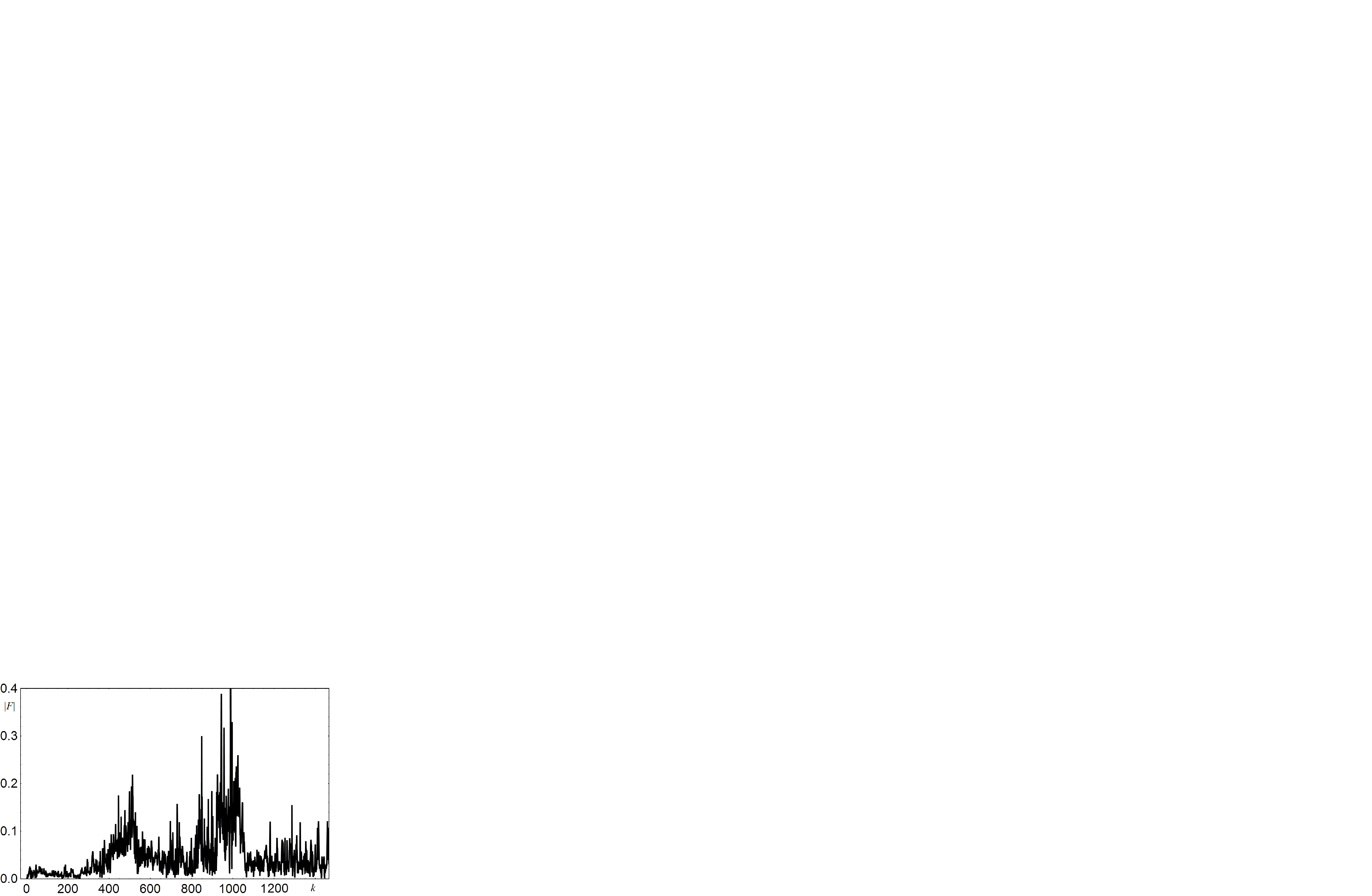}\\
\includegraphics[width=6.5 cm, height=4.5 cm]{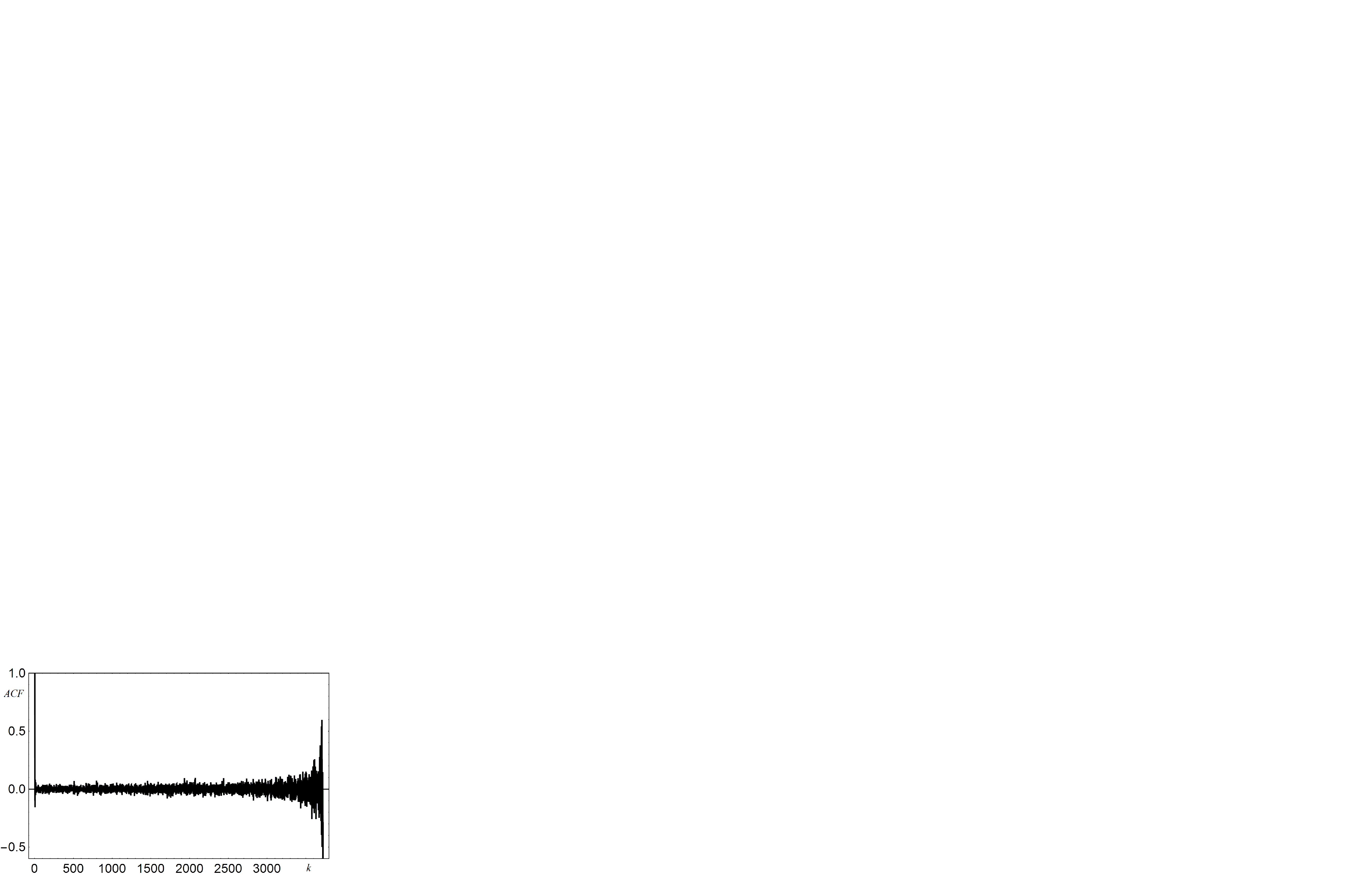}\hspace{0.5cm}
\includegraphics[width=6.5 cm, height=4.5 cm]{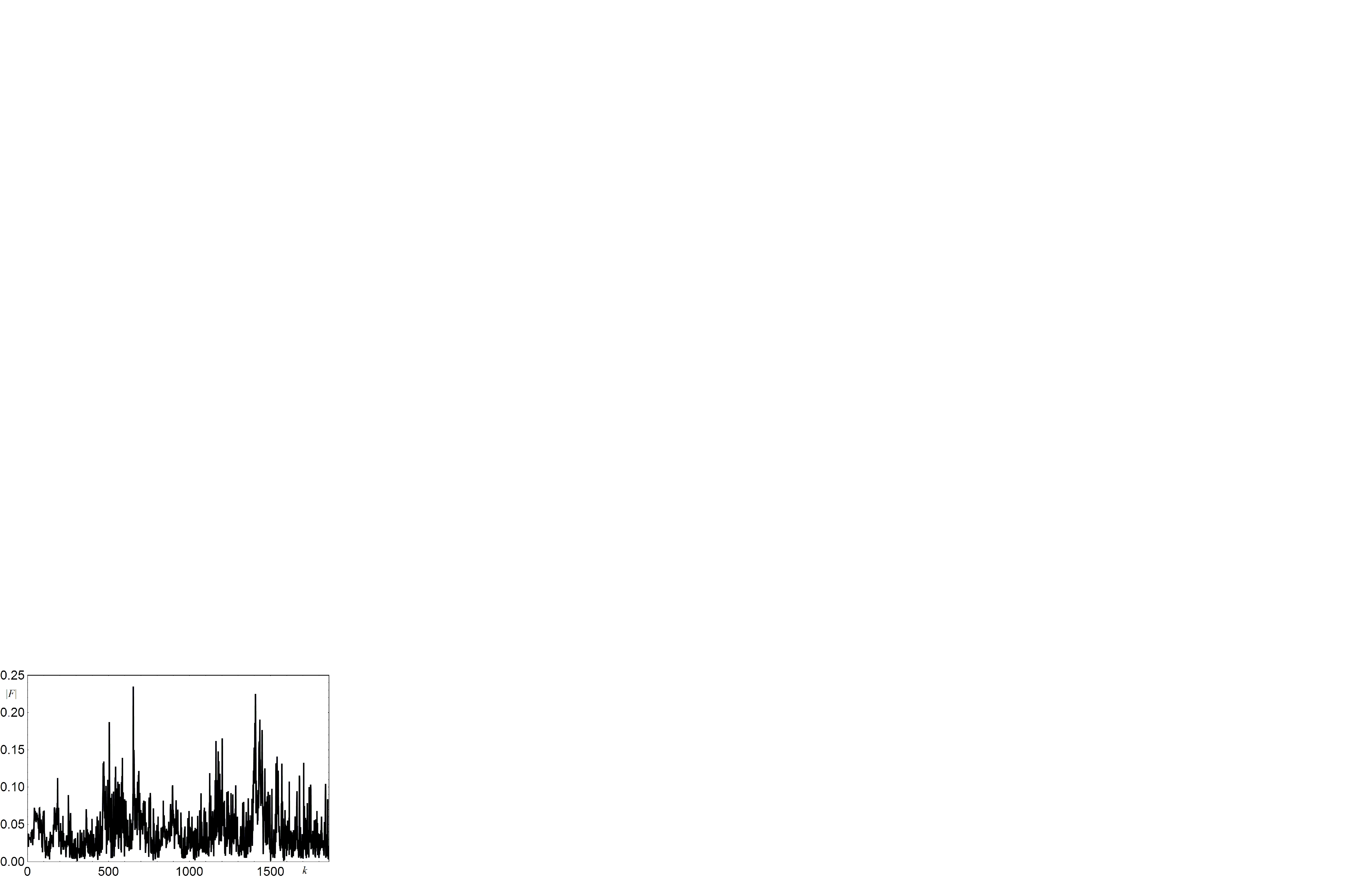}\\
\end{center}
\caption{The autocorralation function (left panels) and its Fourier transformation 
(right panels). Here upper panels correspond to derivations for the numerical model, whereas the lower panels concern the experimental data.} \label{skur:fig10}
\end{figure}

\begin{figure}[bth]
\begin{center}
\includegraphics[width=7.5 cm, height=4.5 cm]{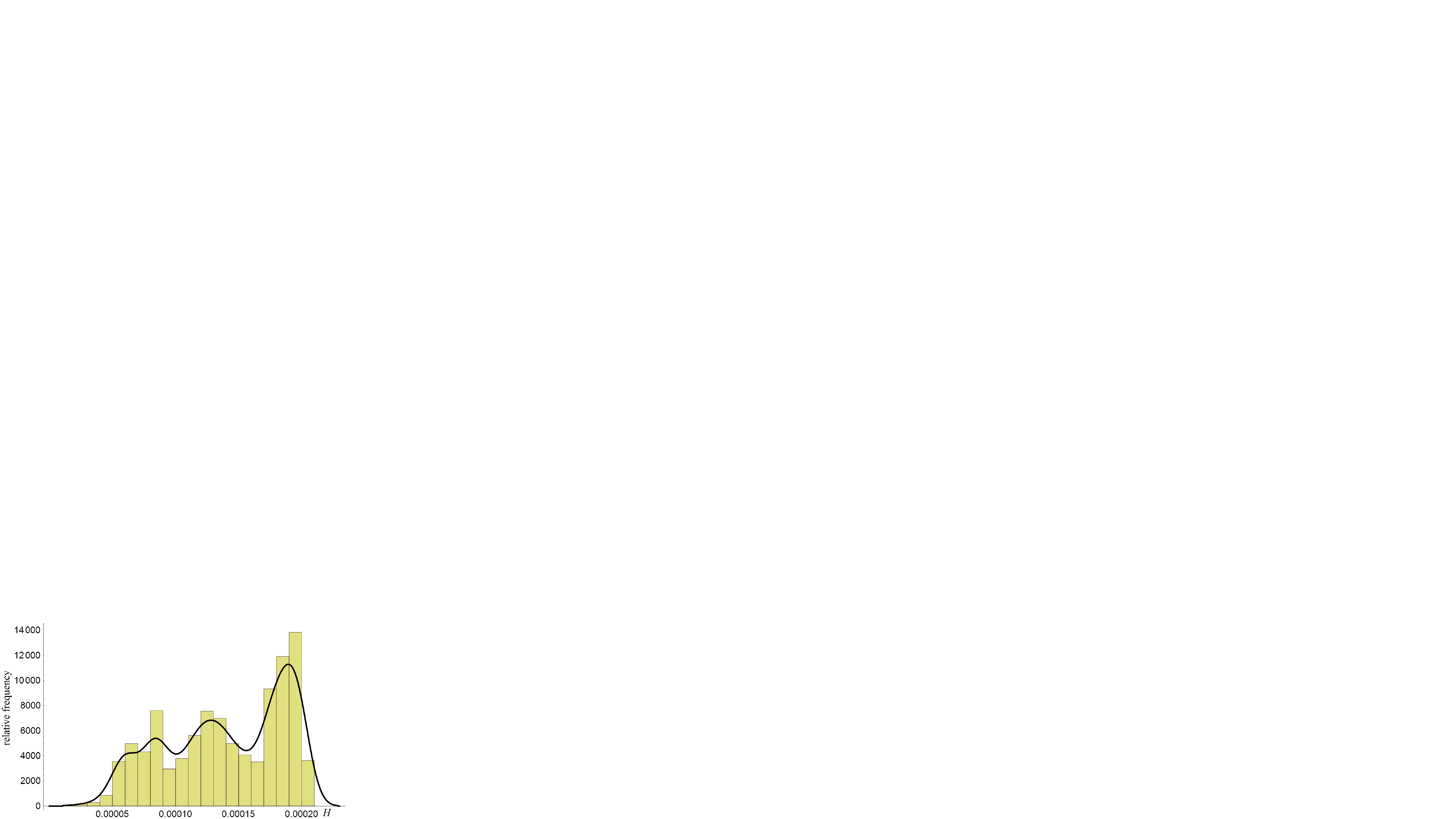}\hspace{0.5cm}
\end{center}
\caption{The relative frequency  histogram for the penetrations of obstacle.
} \label{skur:fig11}
\end{figure}

The important tool for noisy signal analysis is the studies of autocorrelation function (ACF) and its Fourier spectrum. Recall that the autocorrelation concerns the probability to find two impacts at a certain distance \cite{Gerstner2014}. From the signal analysis theory,  ACF defines the similarity of temporally lagged parts of signal. 
Thus, the ACF is defined as follows
\begin{equation}\label{skur:acf}
\mbox{ACF}(k)=\frac{n}{n-k+1}\frac{\sum_{i=1}^{n-k+1}(S_i- <S>)(S_{i-k+1}- <S>)}{\sum_{i=1}^{n}(S_i- <S>)^2},
\end{equation}
where $<S>=\sum_{i=1}^n S_i/n$ is the mean value. 
Using relation (\ref{skur:acf}), we derive ACF  and their Fourier transformations for numerically (upper panel in Fig.\ref{skur:fig10}) and experimentally (lower panel in  Fig.\ref{skur:fig10}) obtained ISI. There are two peaks in the Fourier spectrum for the numerical ISI and two weakly prevailing (due to noise present) extrema in the experimental ISI. These peaks can be associated with the temporal scales in the signal. In particular, these scales can be related with the intervals providing the maxima at the relative frequency histograms (Fig.\ref{skur:fig9}).    
 
 Analyzing the profile of $x$ component in Fig.\ref{skur:fig3}a, the different  height of peaks  is observed. This tells us that the penetrations of the obstacle are different. Consider the distribution of random variable $H_j$ which is the set of maximal values of obstacle penetrations during an impact. To derive the sequence $H_j$, we estimate  the moments of time when the maximum of $x$ is reached, i.e. $x'(t_j)=0$, and $x(t_j)>x_I$. Then  $H_j=x(t_j)-x_I$. Let us form the relative frequency histogram of corresponding values $H_j$. The resulting histogram presented in Fig.~\ref{skur:fig11} possesses one essential maxima relating to the appearance of deepest penetration of obstacle. There are also a couple  of maxima almost twice as small as the main peak. Thus, the distribution  observed is not unimodal but multimodal.

\section{Concluding remarks}

Summarizing, we studied the dynamics of forced impacting oscillator examining the sequences of impacts generated by stop impacts. It turned out that this quite simple mechanical system generates  extremely interesting impact trains especially in the mode of chaotic vibrations. 
In the frequency domain corresponding to the periodic solutions existence allows one to identify the characteristics of regimes and their bifurcations, in particular the grazing bifurcation. When the chaotic regime occurs, the impact sequence is chaotic and is studied from the statistical point of view. At first, the construction of  successive iterations of impact map leads to the discontinuous locus of points corresponding to the beginning of impacts, unlike the classical Poincare section for this model.   Note also that the inter-impact intervals form the stationary non-Poissonian stochastic sequence and obey the bimodal distribution. The similar distribution is distinguished in the experimental data arranged in the proper histogram. 
The multimodal distribution is also revealed  in the sequences  of  obstacle penetration.
Correlation analysis of these sequences  showed the presence of temporal scales.  All these features were identified in both  numerical and experimental investigations.  

\section*
{Acknowledgments}
This work has been supported by the Polish National Science Centre under the grant OPUS 14 No. 2017/27/B/ST8/01330.

\end{document}